\documentclass[11pt,a4paper]{article}
 \usepackage{graphicx}
 \usepackage{afterpage}
 \usepackage{enumerate}
  \usepackage{longtable}

\begin{document}
\small
 \title{\bf Influence of the vertical and horizontal magnetic field inhomogeneity on the Stokes parameters of the magnetically sensitive Fe I line 525.02 nm
}
\author{\bf V.G. Lozitskii$^1$ and V.A. Sheminova$^2$ }
  \date{}

 \maketitle
 \thanks{}
\begin{center}
{ $^1$Astronomical Observatory, Shevchenko National University,\\ Observatornaya 3, Kyiv, 04053, Ukraine\\

      $^2$Main Astronomical Observatory, National Academy of Sciences of Ukraine,\\ Akademika  Zabolotnoho 27,  Kyiv,  03143, Ukraine,  e-mail: shem@mao.kiev.ua
}
\end{center}

\begin{abstract}
{Based on calculations of the Stokes parameters for the Holweger-Mliller model atmosphere, we study sensitivity of the Fe I 525.02 nm line to some kinds of vertical and horizontal magnetic field inhomogeneity. A noticeable asymmetry is shown to appear in the $V$ profile peaks when the vertical gradient is $-0.4$ mT/km, which is typical of some theoretical flux tube models. The asymmetry is most pronounced in a pure longitudinal magnetic field and at a low macroturbulent velocity. A similar effect is observed for the $Q$ profile in nonlongitudinal fields as well. The Fe I 525.02 nm line is sensitive also to subtelescopic fields of mixed polarity like those observed by Stenflo in IR lines. We argue that the Wilson depression in small-scale flux tubes renders strong-field areas invisible at heliocentric angles greater than 60--65$^\circ$, since they are screened by surroundings with weaker magnetic fields.

}
\end{abstract}


\section{Introduction }

A distinguishing feature of solar magnetic fields is their inhomogeneity and fine structure. As of now, many physical parameters of fine-structure elements, such as field intensity, minimum size, vertical gradient, magnetic fields in the interspace between elements, and some others, have been studied insufficiently. This has to do with small dimensions of the finest elements in solar magnetic fields, which are not larger than $10^2$ km, while a typical resolution in direct ground-based observations is usually about $10^3$ km [7,23]. These finest elements are spatially unresolvable, and we call them subtelescopic elements. The dimensions of such elements being by about an order of magnitude smaller than the effective aperture diameter, it is likely that they do not cover the whole aperture area but its part only (an exception may be solar spots, where these elements may be immediately adjacent to one another [10]). The portion of the aperture area covered by small-scale elements is usually called the filling factor $ \alpha$. Such structures can be modeled as a flux tube with a sufficiently high field strength, about 0.1--0.2 T [18,23,28,29] or higher [4,5,9]. The filling factor in quiet regions (i.e., in the network) is usually 3--5\%, while it is 10--15\% in faculae [21,27] and can be as high as 40\% in flares [20].

As we usually have $\alpha < 1$, special indirect methods, for instance, the line-ratio method [23], rather than direct ones have to be used in studying small-scale flux tubes. The results depend on initial assumptions, for instance, on the magnetic field structure inside the flux tube, plasma motions in tubes, background field parameters. This calls for preliminary investigations of the effect of some factors on measurement results. Among other things, it is of importance to find how magnetic field inhomogeneities affect the profiles of magnetically sensitive lines. We intend in the present paper to elucidate this effect as to the vertical gradient of magnetic field in flux tubes and the background fields of different intensity and polarity.

A vertical magnetic field gradient develops inevitably owing to gas pressure drop with height in the atmosphere. Flux tubes may be expected to spread upwards and to form a continuous quasi-uniform field of a magnetic blanket type in the temperature minimum region [15]. There is observational evidence that the magnetic field structure in the temperature minimum zone may differ at different locations on the Sun: it may be either a quasi-uniform field or small-scale tubes [7]. If a blanket-type quasi-uniform field does really form in this zone, the vertical gradient $dB/dh$ of magnetic field in flux tubes must be negative and no more than 0.5 mT/km in absolute value. This result was obtained by the line-ratio method using data for the lines Fe I $\lambda$ 525.35 nm and Mg I $\lambda$ 518.4 nm. Approximately the same vertical gradient was obtained in [18]. Theoretical model [31] yields $-$0.3 mT/km for the gradient at a height of 200 km. Even greater gradients are possible in a force-free model [11], from $-$1 to $-$10 mT/km, depending not only on the external gas pressure but on the magnetic field topology and strength on tube axes as well. Conceivably such a magnetic field structure might occur sometimes in active regions and in flares. It was found in a flare of importance 2 that the split in emission peaks in the $I\pm V$ profiles of Fe I lines suggested a vertical gradient of $-$1 mT/km in flux tubes, this value agreeing well with force-free model [11]. On the whole it is not clear, however, to what extent the topology and height variations of magnetic fields in flux tubes pictured in the above-mentioned studies are adequate to the actual situation.

The magnetic field intensity in the interspace between flux tubes is also unknown. It was postulated in the early empirical flux tube models based on the data about the longitudinal Zeeman effect that the interspace between flux tubes is filled by a nonmagnetic plasma [23]. It was found later, however, from the studies of the Hanle effect that diffuse background magnetic fields also contained a large hidden magnetic flux [24]. Recent measurements of nonspot fields with the use of magnetically sensitive lines in the infrared ($\lambda \approx 1.6$ microns) reveal that the interspace between usual flux tubes with high intensities contains discrete magnetic elements of different polarities, with typical strengths of several tens of mT [25]. A similar magnetic field structure was noted earlier in the photosphere in an active region and in a flare [6]. Although it follows from [6,25] that components with a strong or a moderate field would suffice to account for $I$ and $V$ profiles of spectral lines, the existence of weaker background fields cannot be excluded altogether. This is suggested by measurements made with a high (1--2$^{\prime\prime}$) spatial resolution with the line-ratio method in quiet areas [7].

\section{Computation of the theoretical Stokes profiles}

The transfer equations for the radiation polarized in a magnetic field, which are often called the Unno--Rachkovskii equations, are a set of four first-order differential equations. The Stokes parameter profiles for the emergent flux can be obtained using the fifth-order Runge--Kutta--Fehlberg method and the boundary conditions in accordance with [19]. This has been realized in the SPANSATM program package. The algorithm and the program potentialities are described in [8,12,13]. The only essential constraint in the algorithm is the LTE condition. Nevertheless, the NLTE effects can be allowed for empirically using coefficients of deviation from LTE.

The Stokes parameters were calculated for the line Fe I $\lambda$ 525.02 nm, which is often used in the measurements of solar magnetic fields. This line belongs to the first multiplet of iron, its low excitation potential is $EP = 0.12$~eV, the effective Lande factor is $g_{\rm eff} = 3.0$. The calculations were done with the following input data: the HOLMU model atmosphere [17], microturbulent velocity $v_{\rm micro} = 0.8$~km/s, macroturbulent velocity $v_{\rm macro}$ from 0 to 3 km/s, damping constant $\Gamma = 1.5\Gamma_{\rm WdW}$, iron abundance $A = 7.64$, oscillator strength $\log gf= -4.89$ [2]. The Stokes profiles $I,~Q,~U,~V$ were obtained in the relative depression units: $1 - I/I_c,~-Q/I_c,~-U/I_c,~-V/I_c$, where $I_c$ is the continuum radiation intensity. For simplicity, we omit the minus sign before $Q/I_c, ~U/I_c$ and $V/I_c$ in the subsequent text.

\section{Height gradient of the magnetic field}

We studied the Stokes profiles for four magnetic field models (Fig. 1). The field in Model 1 decreases monotonically by a linear law from the strength $B= 0.25$~T at the height $h = 0$ to $B = 0$ at $h = 600$~km. Thus the magnetic field in this model exists along the whole height in the atmosphere and has the constant gradient $dB/dh = -0.42$~mT/km. This gradient is close to the theoretical one in model [31], where $dB/dh$ is $-$0.40 mT/km at $h = 0$ and about $-$0.25 mT/km at $h = 300$ km. The gradients in model [22] are about the same.

The run of the magnetic field strength in Model 2 is the reverse of that in Model 1. The $dB/dh$ value is also constant along the height and is 0.42 mT/km. This model might appear at first glance to be abstract and unrealistic, but it may prove to have some relevance to reality under certain special conditions (for instance, when horizontal electric currents local in height are present [1]).

Models 3 and 4 are characterized by a nonmonotonic run of magnetic strength with height. The gradient in Model 3 is +0.42 mT/km in the height range 0--300 km and -0.42 mT/km in the range from 300 to 600 km. The absolute gradients are twice as high in Model 4, and their sign also changes abruptly at a height of 300 km. These models are of interest, since they demonstrate the sensitivity of the Stokes profiles to local rises of magnetic field strength. Such effects have been suspected to exist in flares in particular [3]. As to the photosphere, the line Fe I $\lambda$ 525.02 nm is just an appropriate indicator for it, the line's contribution function covering almost the whole height of the photosphere, with the effective height of line formation $h = 330$~km [14].
 \begin{figure}[h!b]
 \centerline{
\includegraphics [scale=1.]{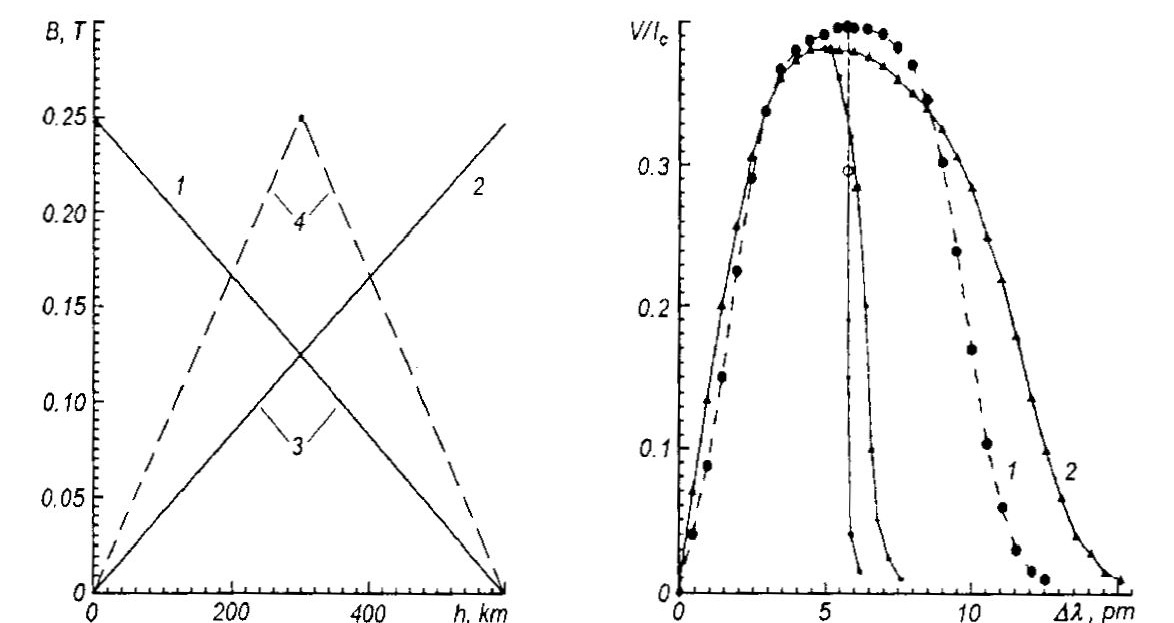}
     }
  \caption {\small
The magnetic field in Model 1 decreases monotonically, in Model 2 is the reverse of that in Model 1, in Models 3 and 4 are characterized by a nonmonotonic run of magnetic strength with height.
}
 \caption {\small
$V/I_c$-parameter profiles for the homogeneous magnetic field $B = 150$~mT (curve 1, circles) and for Model 1 (curve 2, triangles) at $\gamma = 0^\circ$ and $v_{\rm macro} = 0$ (see the text).
}
\end{figure}

Model 1 exhibits an appreciable asymmetry of the peaks in the Stokes $V$ parameter (Fig. 2), while a similar case of a homogeneous field with $B =$~const~=~ 0.15 T gives no such asymmetry either with a longitudinal field $\gamma = 0^\circ$ or at large angles of inclination of field lines. This asymmetry is typical, however, just in pure longitudinal fields because it produces a deformation only in the peak of the V-parameter wave, making an impression that it is double. When this effect is analyzed by studying the bisectors of V profiles, we find that the bisectors bend markedly toward the line centers near the profile tops only at $\gamma = 0^\circ$, they remain rectilinear (as in the case $B =$~const) on other profile sections, except in far line wings, of course. At large angles $\gamma$, for instance 70--80$^\circ$, the bisectors are distorted along the whole $V$-profile height.

To have a quantitative characteristic for the deformation of $V$-profile peaks, we consider the asymmetry parameter
\begin{equation}
  A=\frac{a_{1.0}- a_{0.75}}{a_{1.0}+ a_{0.75}},
\end{equation}
where $a_{1.0}$ and $a_{0.75}$  are the distances from the line center to those bisector sections which correspond to the $V$-profile peak and to a level of 3/4 of its maximum height. The corresponding bisector sections are shown by larger marks in Fig. 2 and subsequent figures where profile bisectors are depicted. The subscripts $V$ and $Q$ with the parameter $A$ denote that this parameter refers to the $V$ and $Q$ Stokes profiles, respectively.

In Model 1 of the magnetic field ($dB/dh < 0$) $A_V < 0$ and $A_Q < 0$, and these quantities depend essentially on both the angle $\gamma$ of inclination of field lines to the line of sight and the macroturbulent velocity $v_{\rm macro}$ (Fig. 3). Let us examine this dependence in more detail (Fig. 4).

When the magnetic field is purely longitudinal ($\gamma = 0^\circ$), the least value of $A_V$ (-8.0\%) is attained at $v_{\rm macro}= 0$. When the velocity grows, $A_V$ grows monotonically to about $-2$\% at $v_{\rm macro} = 3$ km/s. The absolute value of $A_V$ diminishes also when $\gamma$  grows (curve 2 in Fig. 4). In particular, the absolute values of $A_V$ are at least half as large at $\gamma = 75^\circ$ than at $\gamma = 0^\circ$, their change being nonmonotonic; they attain their absolute minimum ($A_V = 0$) at $v_{\rm macro} =1.5$ km/s. When we compare these changes in $A_V$ with the corresponding changes for a homogeneous field, we find that $A_V$ is zero at $v_{\rm macro}$ for 0 to 1.5 km/s and it is independent of the inclination of field lines. At $v_{\rm macro} > 1.5$ km/s $A_V$ differs from zero but no more than by 2\% (with $\gamma = 0^\circ$ and $v_{\rm macro} = 3$ km/s).
 \begin{figure}[h!b]
 \centerline{
\includegraphics [scale=1.]{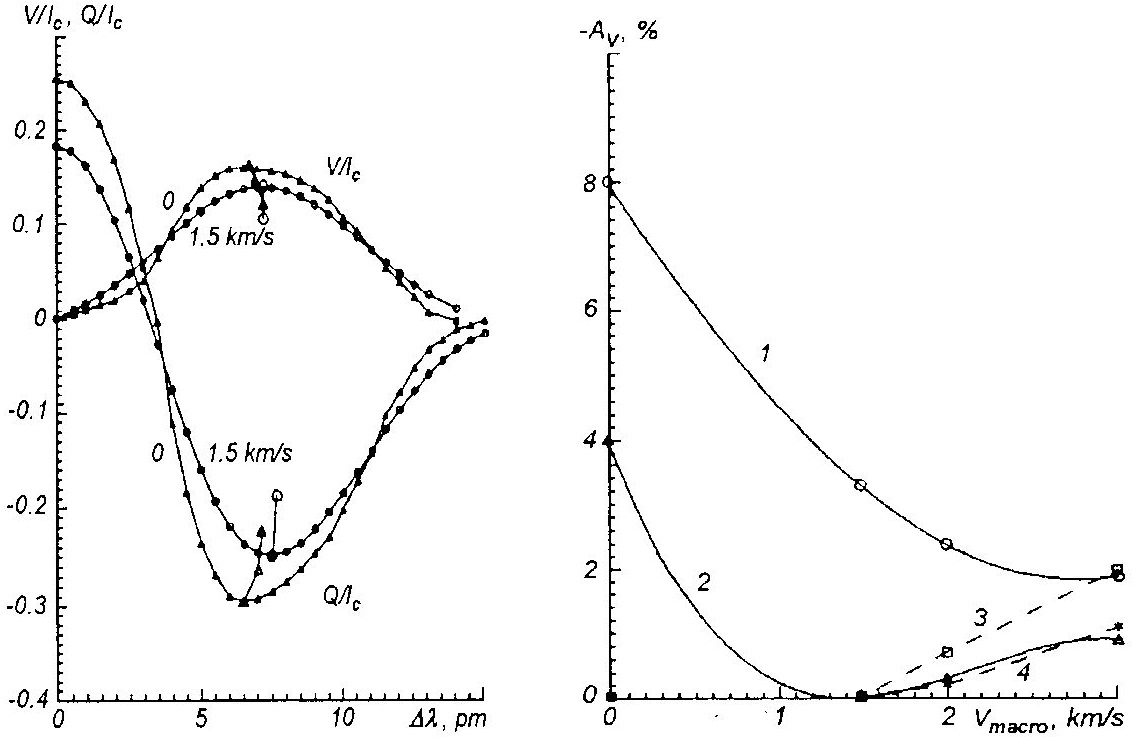}
     }
  \caption {\small
Profiles of the parameters $V/I_c$ and $Q/I_c$ for Model 1 at $\gamma = 75^\circ$ and $v_{\rm macro} = 0$ and 1.5 km/s. In each profile those bisector sections are shown which were used for the determination of the asymmetry parameter $A_V$ (see the text).
}
\caption {\small
Asymmetry parameter $A_V$ vs. macroturbulent velocity $v_{\rm macro}$: 1) Model 1, $\gamma = 0^\circ$; 2) Model 1, $\gamma = 75^\circ$; 3) homogeneous field, B = 150 mT, $\gamma = 0^\circ$; 4) the same, $\gamma = 75^\circ$.
}
\end{figure}

It is important for the diagnostics of the magnetic field height gradient that a homogeneous field produces rather fine asymmetry effects which, in addition, become evident only at $v_{\rm macro}> 1.5$ km/s. In the actual undisturbed solar atmosphere just $v_{\rm macro} \approx 1.5$ km/s is the case in a wide height range in the photosphere [2]. Besides, the turbulent velocity in small-scale flux tubes is smaller than in the undisturbed photosphere [7]. This permits us to hope that it is quite possible to diagnose the height gradient in the actual small-scale magnetic fields on the Sun using the asymmetry parameter $A_V$.

Model 2 exhibits an insignificant positive asymmetry ($A_V$ = 0.5\% at $v_{\rm macro} = 0$; Fig. 5). However, one can obtain the corresponding effect of the same magnitude as in Model 1 if $I\pm V$ profiles rather than $V$ profiles are treated. The asymmetry is also positive in Models 3 and 4, in particular we have $A_V =4.6$\% for $\gamma = 0^\circ$ and $v_{\rm macro} = 0$ in Model 4. The bisectors in two latter models do not differ qualitatively in their shape from those in Model 2. Thus the alternating height gradient does not cause any specific deformations in the Stokes parameter bisectors: they appear the same as in a monotonic variation of the field with height. This suggests that the diagnostics of magnetic fields with an alternating height gradient should be based on several lines of the Fe I $\lambda$ 525.0 nm type rather than on a single line.

Now we compare the above data with observations. The $V$ profiles given in [26] point to $A_V$ = $-2$\% in a faint facula and $-1$\% in a bright facula. Further reduction of the observations analyzed in [20] revealed that $A_V$ ranged from 0 (flare maximum) to $-4$\% (flash phase) in a class 2 flare. The latter value (greater in magnitude) points to a gradient of about $-0.4$ mT/km if we assume that $\gamma = 0^\circ$ and $v_{\rm macro} =1.5$ km/s. Thus these observational data are testimony to the negative gradient of the magnetic field in the solar atmosphere, though we cannot determine its exact value here, as $\gamma$ and $v_{\rm macro}$ are not specified in [20,26]. To define concretely $\gamma$, one needs the $Q$ and $U$ profiles, the more so as these profiles are sensitive to the vertical inhomogeneity of the magnetic field. In particular, we have $A_Q = -5.2$\% at $\gamma = 75^\circ$ when $v_{\rm macro} = 0$ and $A_Q = -1.4$\% when $v_{\rm macro} =1.5$ km/s (for Model 1). Using data on the both circular and linear polarization, we may determine $dB/dh$ more exactly. But, on the other hand, there still remains the problem of $v_{\rm macro}$, which is by no means easy to solve for the quiet photosphere [2]. Below we discuss the lines of attack on this problem as applied to small-scale magnetic structures.
 \begin{figure*}[h!b]
 \centerline{
\includegraphics [scale=1.0]{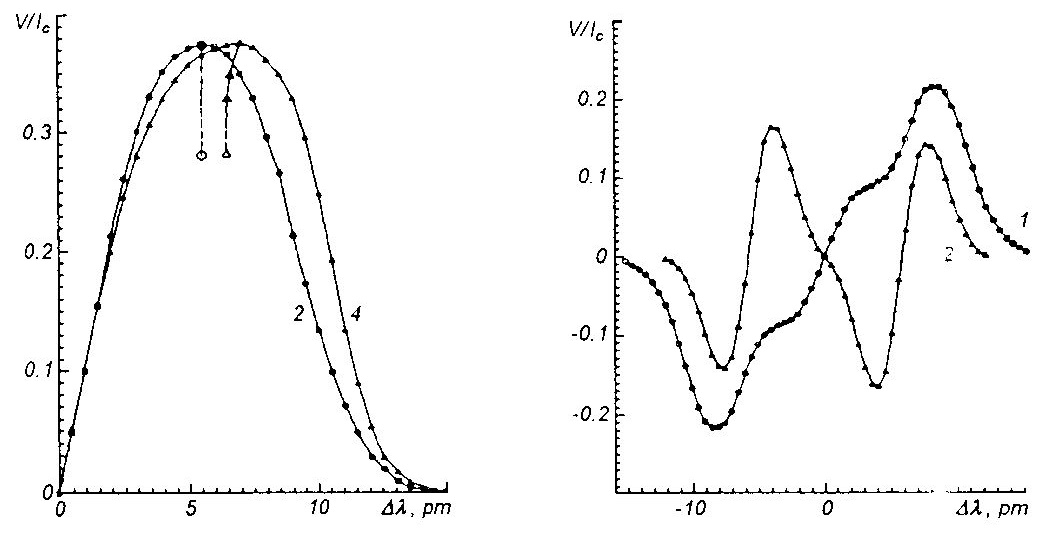}
    }
  \caption {\small
$V/I_c$-parameter profiles calculated for the magnetic fields of Model 2 and Model 4 with  $\gamma = 0^\circ$, $v_{\rm macro} = 0$.
}
 \caption {\small
 $V/I_c$ profiles calculated for magnetic fields of mixed polarity: 1)	$B_1 = 150$ mT, $B_2 =-50$ mT, $\gamma_1 = \gamma_2 = 0^\circ$, $\alpha_1 =\alpha_2$, $v_{\rm macro}= 1.5$~km/s; 2)	$B_1 = 150$ mT, $B_2 =-50$ mT, $\gamma_1 = 75^\circ$, $\gamma_2 = 45^\circ$, $\alpha_1 =\alpha_2$, $v_{\rm macro} = 0$.
}
\end{figure*}

\section{Background field }

We have already noted that new observational data [25] point to the presence of discrete areas with magnetic fields of tens of mT with different polarity in the interspace between common flux tubes. It is pointed out in [25] that such fields could be detected owing to the use of magnetically sensitive lines in the IR spectrum ($\lambda = 1.6$ $\mu$m) which are about three times more sensitive to the magnetic field than lines in the visible spectrum because the Zeeman splitting in them is proportional to $g\lambda^2$. The Stokes $V$ profiles for two Fe I lines with wavelengths 1.5648 $\mu$m and 1.5653 $\mu$m and Lande factors 3.0 and 1.5 were found to be unusual in shape: they have two positive and two negative peaks, while they must have one peak for each sign in a homogeneous field. This result is explained in [25] by the existence of oppositely directed fields, with intensities of 170 mT and $-105$ mT in this specific case. The stronger positive field is attributed to common flux tubes and the fainter negative field to background features. It is pointed out also that the Stokes synthetic $V$ profiles calculated with these intensities for lines in the visible region have a quite normal shape -- they have one peak for each sign.

The latter statement seems to be controversial. Some lines in the visible region, such as the Fe I triplets $\lambda\lambda$ 525.02, 617.33, and 630.25 nm, are partially splitted already ($\Delta\lambda_H\approx\Delta\lambda_D$) at intensities of 100--150 mT and may also reflect, therefore, the existence of fields with such intensities. Nevertheless, numerical calculations are necessary for more reliable inferences. With this in mind, we calculated theoretical profiles for the line Fe I $\lambda$ 525.02 nm for the cases when fields of different intensities and signs are mixed. Assume that we have two magnetic components with the intensities $H_1$ and $H_2$ and filling factors $\alpha_1$  and $\alpha_2$ within the entrance aperture of our instrument. It
is easily shown that the observed $V/I_c$ profile can be calculated as
\begin{equation}
  \frac{V}{I_c} = (1-\beta)\left[\alpha_1 \frac{I_{c,1}}{I_{c}}\frac{V_1}{I_{c,1}}+\alpha_2 \frac{I_{c,2}}{I_{c}}\frac{V_2}{I_{c,2}}\right  ]+ \beta\frac{V_s}{I_c},
 \end{equation}
where $V_1,~ V_2$, and $V_s$ are the $V$-parameter profiles referring to the first magnetic component, to the second one, and to the scattered light, respectively; $\beta$ is the fraction of scattered light; the ratios $I_{c,1}/I_c$ and $I_{c,2}/I_c$ are relative brightnesses in the continuum for two magnetic components. To reveal the major effects, we assumed $I_{c,1} \approx I_{c,2} \approx I_{c}$. and $\beta = 0$. Then we have
\begin{equation}
  \frac{V}{I_c} \simeq \alpha_1 \frac{V_1}{I_c} +\alpha_2 \frac{V_2}{I_c}. \end{equation}

Figure 6 gives the results of calculations with (3). The line Fe I $\lambda$ 525.0 nm as an indicator of such fields was found to be quite adequate for their diagnostics, though it is inferior to the lines in the infrared. The $\lambda$ 525.0 nm line has a markedly distorted $V/I_c$ profile at intensities of 150 and $-50$ mT, $\gamma = 0^\circ$, and $v_{\rm macro}= 1.5$ km/s. It does not show two positive and two negative peaks at these parameters as the IR lines do, but the peaks appear when $v_{\rm macro} = 0$ and $\gamma_1  \gamma_2 \neq 0^\circ$. Such a picture is observed, in particular, for $B_1 = 150$ mT, $B_2 = -50$ mT, $\gamma =75^\circ$, $\gamma =45^\circ$, and $\alpha_1 = \alpha_2$. The principal and secondary peak amplitudes are virtually the same in this case. A qualitatively similar result can be obtained for $\gamma_1 =\gamma_2 = 0^\circ$ also but in the assumption that $\alpha_1\neq \alpha_2$, for instance for $B_1= 170$ mT, $B_2 = -100$ mT, $\alpha_2 = 1.7\alpha_1$, i.e., when the absolute fluxes are equal in the magnetic fields of both signs.

\section{ Discussion}

The above calculation results suggest that the Fe I $\lambda$ 525.02 nm line is quite adequate for the diagnostics of the both vertical and horizontal inhomogeneity of the magnetic field in small-scale flux tubes and of background fields as well. The flux tubes being predominantly vertical structures in the atmosphere [25] and the most significant deformations of the Stokes parameters occurring in the longitudinal field owing to the height gradient, preference should be given to the observations of the central parts of the solar disk when the vertical inhomogeneity is studied. When the horizontal inhomogeneity is studied, areas more distant from the center of the disk may be studied, though a certain limit seems to exist here due to Wilson's depression in flux tubes (the continuum and spectral lines must be formed at lower levels in the photosphere). The depression makes the observed solar surface ragged with numerous craters and cirques. Craters and
cones develop on the Sun due to a difference of the gas pressure in flux tubes and outside them. For a flux tube to be in equilibrium, it is necessary that
\begin{equation}
P+\frac{B^2}{8\pi}=P_{\rm ex},
\end{equation}
where $P$ and $P_{\rm ex}$ are the gas pressures inside and outside the flux tube, respectively, $B$ is the magnetic field intensity in the tube (we ignore the field intensity outside the tube). It is obvious that $P < P_{\rm ex}$, i.e., the gas inside the tube is more transparent, and therefore the bottom of magnetic cone seems to lie lower than the surrounding photosphere.

A strong magnetic field in a magnetic tube can be detected by the Zeeman effect only in the case when the magnetic cone bottom is observed. When the line of sight is inclined to the solar surface (e.g., near the limb) we observe cone walls rather than the bottom, and then the effect of strong fields must disappear. Therefore $\gamma$ cannot be too close to $90^\circ$ for the flux tubes with $\alpha < 1$ spread over the surface.

The Fourier transform spectrometer technique provides valuable data [27] which bear witness to the existence of screening of strong fields in flux tubes due to Wilson's depression. Drastic variations were found to occur in some $V$-peak parameters for the Fe I lines $\lambda\lambda$ 524.71 and 525.02 nm (namely, in their splitting and asymmetry) near the heliocentric angle $\theta$ for which $\mu = \cos\theta = 0.4$--0.5 (see Figs 4 and 6 in [27]).

When we have a tube cone of diameter $d$ and height $z_W$, we can see the cone bottom only if $\mu >\mu_{\rm cr}$, where  $\mu_{\rm cr}$ is the critical value $\mu$ found from the relation
\begin{equation}
  d=z_W(\mu_{\rm cr}^{-2}-1)^{1/2}.
\end{equation}

If we assume that $\mu_{\rm cr} \approx 0.45$ based on the results of [26], we have $\theta_{\rm cr} \approx\gamma_{\rm cr} \approx 63.5^\circ$ and $d/z_W \approx 2$. In this case photospheric flux tubes should not be observed at large heliocentric angles (e.g., 70--80$^\circ$). This provides one explanation more for the center-to-limb effect in magnetographic observations [16] of longitudinal magnetic fields $H_\parallel$ in different spectral lines $\lambda_i$, and $\lambda_j$ (the essence of the effect is that the ratios $ H_\parallel(\lambda_i)/H_\parallel(\lambda_j)$ tend to unity when approaching the limb, though they may differ from unity by a factor of 2 or 3). This effect arises not only because the magnetic field weakens with height or because the temperature differences in the field tubes -- background gaps are erased with height, but largely because at large heliocentric angles, areas with strong field. Obviously, the phenomenon of screening of strong fields due to the Wilson depression can be used as a tool for studying the height and surface structure of flux tubes.

In conclusion, let us briefly focus on the magnitude of the $v_{\rm macro}$ in solar magnetic formations, on which, as was shown above, the degree of certainty of conclusions about inhomogeneous magnetic fields depends. Apparently, this problem can be solved using several specially adapted spectral lines, which have the same depth of formation and temperature sensitivity, but different factors of Lande. For example, it is possible to use the Fe I 524.7 and 525.0 nm lines with Lande factors 2.0 and 3.0, respectively, which were previously recommended for measurements by the method of the ratio of lines [23]. This can make it possible to separate the contributions from the magnetic and nonmagnetic expansion of the Stokes profiles and thereby eliminate the influence of the $v_{\rm macro}$ on the measurement results.

\vspace {0.3cm }
The authors are grateful to R.I. Kostyk and the reviewer for valuable comments.
\vspace {0.3cm }

\end{document}